\documentclass{appolb}
\usepackage{graphicx}
\usepackage{subcaption}
\usepackage{amsmath}
\usepackage{braket}

\begin{document}
	\eqsec  
	\title{Structure of hybrid static potential flux tubes in SU(2) lattice Yang-Mills theory%
		\thanks{Presented at Excited QCD 2018}%
	}
	\author{Lasse M{\"u}ller, Marc Wagner
		\address{Goethe-Universit\"at Frankfurt am Main, Institut f\"ur Theoretische Physik, Max-von-Laue-Stra{\ss}e 1, D-60438 Frankfurt am Main, Germany}}
	\maketitle
	
	\begin{abstract}
		We study the structure of the hybrid static potential flux tube in the $\Pi_u$ sector in SU(2) lattice Yang-Mills theory. To this end, we compute the squares of the chromoelectric and chromomagnetic field strengths in the presence of a static quark-antiquark pair. We show clear evidence that the gluon distribution is significantly different compared to that of the ordinary static potential with quantum numbers $\Sigma_g^+$.
	\end{abstract}
	

\section{Introduction}

Hybrid mesons are systems composed of a quark-antiquark pair and excited gluons, where the latter allow quantum numbers $J^{PC}$ not possible in the quark model. Such hybrid mesons are subject of current investigations both in experimental and theoretical physics. On the theoretical side lattice field theory is particularly suited to study hybrid mesons with two very heavy quarks, i.e.\ a quark-antiquark pair in the static limit. While there are several existing computations of hybrid static potentials (cf.\ e.g.\ \cite{Michael:1998tr,Bali:2000vr,Juge:2002br,Bali:2003jq,Reisinger:2017btr} and references therein), the corresponding flux tubes have been studied for the first time only recently \cite{Bicudo:2018yhk} (for completeness we also mention an earlier paper \cite{Cardoso:2009kz} by the same authors, where hybrid static potential flux tubes have been studied using not only static quarks, but also static gluons).

In the following we present the status of our ongoing computations of hybrid static potential flux tubes. We use methods quite similar to those used in \cite{Bicudo:2018yhk}. We find, however, results, which are significantly different from the results presented in \cite{Bicudo:2018yhk} as we will detail at the end of section~\ref{SEC004}.


\section{Hybrid static potentials and flux tubes}

A hybrid meson is a meson, where not only the quark $Q$ and the antiquark $\bar{Q}$, but also the gluons contribute to the quantum numbers $J^{P C}$. When studying hybrid mesons in the limit of infinitely heavy quarks, i.e.\ static quarks $Q$ and $\bar{Q}$, symmetries are somewhat different, resulting in the following three quantum numbers (cf.\ e.g.\ \cite{Juge:2002br,Reisinger:2017btr} for a detailed discussion).
\begin{itemize}
\item The absolute value of angular momentum with respect to the $Q \bar{Q}$ separation axis (here the $z$ axis): $\Lambda \in \{ \Sigma \doteq 0 , \Pi \doteq 1 , \Delta \doteq 2 \}$.

\item The eigenvalue of parity combined with charge conjugation $P \circ C$: $\eta \in \{ g \doteq + , u \doteq - \}$.

\item The eigenvalue of the spatial reflection along an axis perpendicular to the $Q \bar{Q}$ separation axis (here the $x$ axis): $\epsilon \in \left\{ +, -\right\}$.
\end{itemize}
Computing the energy of the groundstate in a given $\Lambda_\eta^\epsilon$ sector as a function of the $Q \bar{Q}$ separation $r$ yields a static potential denoted by $V_{\Lambda_\eta^\epsilon}(r)$. While $V_{\Sigma_g^+}(r)$ is the well known ordinary static potential, all other potentials $V_{\Lambda_\eta^\epsilon}(r)$ exhibit significantly larger energies and are referred to as hybrid static potentials (for a summary plot cf.\ e.g.\ \cite{Juge:2002br}, Fig.\ 2).

	\begin{figure}[b]
		\centerline{%
			\includegraphics[width=5.5cm]{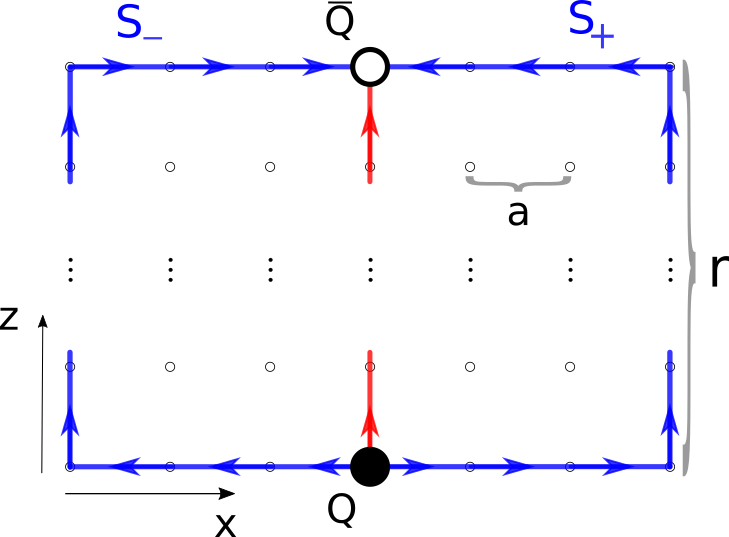}}
		\caption{Spatial part of a Wilson loop suited to study the $\Pi_u$ sector.}
		\label{Fig:Wilson_loop_Pi_u}
	\end{figure}

$V_{\Sigma_g^+}(r)$ can be computed from standard rectangular Wilson loops. To access $\Lambda_\eta^\epsilon$ sectors different from $\Sigma_g^+$, one has to replace the two spatial sides of the Wilson loops by more complicated structures. We focus on the $\Pi_u$ hybrid static potential \footnote{We omit the quantum number $\epsilon$, since hybrid static potentials with $\Lambda \geq 1$ are degenerate with respect to $\epsilon$.} by considering Wilson loop-like correlation functions with spatial parts as shown in Fig.\ \ref{Fig:Wilson_loop_Pi_u}.
With quark and antiquark positions at $\textbf{0}$ and $\mathbf{r} = r \mathbf{e}_z$, respectively, the resulting Wilson loop reads $W_{\Pi_u} = \mathcal{O}(0, \textbf{0}) S_t(0, \mathbf{r}) O^\dagger(t, \textbf{0}) S_t^\dagger (0, \textbf{0})$, where $S_t$ denotes a straight path of $t/a$ links in temporal direction, $\mathcal{O}(t,\textbf{r}) = S_+(t, \textbf{r}) - S_-(t, \textbf{r})$ and $S_+$ and $S_-$ as defined in Fig.\ \ref{Fig:Wilson_loop_Pi_u}.

The main goal of our work is to study hybrid static potential flux tubes by computing expectation values of squares of chromoelectric and chromomagnetic field strength components, $E_j^2$ and $B_j^2$ ($j = x,y,z$ and the square implies a sum over color indices). We have carried out such computations in the presence of a quark-antiquark pair in the $\Sigma_g^+$ and $\Pi_u$ sectors as functions of the spatial coordinate $\textbf{x} = (x,y,z)$ with respect to the vacuum using the equations
	\begin{align}
		\braket{E_j(\textbf{x})^2}_{Q\bar{Q}} - \braket{E_j^2}_{\textrm{vac}}
		&\propto \bigg(\frac{\braket{W \cdot P_{0j}(t/2, \textbf{x})}}{\braket{W}} - \braket{P_{0j}} \bigg) \label{Eqn: E_squared}\\
		\braket{B_j(\textbf{x})^2}_{Q\bar{Q}} - \braket{B_j^2}_{\textrm{vac}}
		&\propto \bigg(\braket{P_{kl}} - \frac{\braket{W \cdot P_{kl}(t/2, \textbf{x})}}{\braket{W}} \bigg), \label{Eqn: B_squared}
	\end{align}
where $P_{0j}$ denotes a spacetime plaquette and $P_{kl}$ a spatial plaquette in the plane perpendicular to the $j$ direction.


	\begin{figure}[b]
		\centerline{%
			\includegraphics[width=10cm]{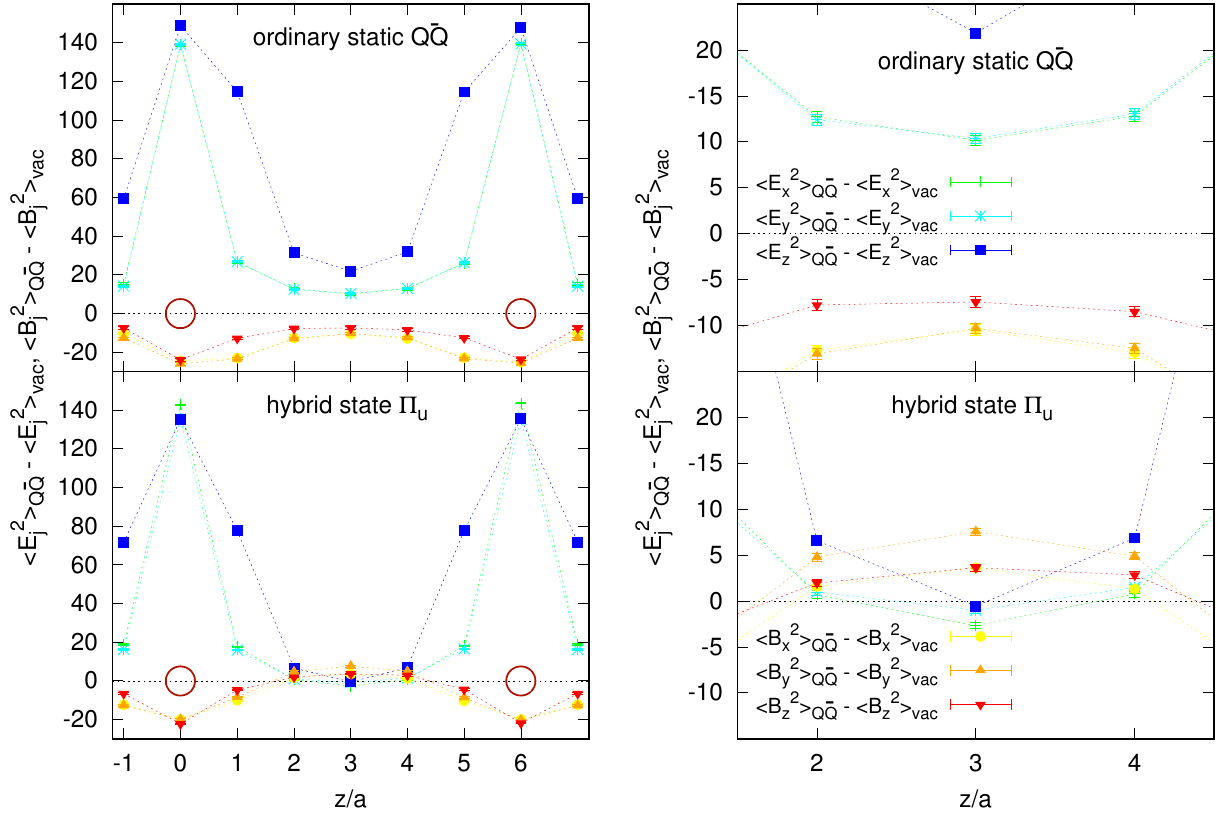}}
		\caption{$\braket{E_j(\textbf{x})^2}_{Q\bar{Q}} - \braket{E_j^2}_{\textrm{vac}}$ and $\braket{B_j(\textbf{x})^2}_{Q\bar{Q}} - \braket{B_j^2}_{\textrm{vac}}$ on the $Q \bar{Q}$ separation axis for separation $r = 6a$. Top: ordinary static potential $\Sigma_g^+$. Bottom: hybrid static potential $\Pi_u$. Plots on the right are zoomed versions of the plots on the left. Red spheres indicate $Q \bar{Q}$ positions at $z = 0$ and $z = 6a$.}
		\label{Fig:separation_axis}
	\end{figure}

\section{Lattice setup}
	
At the moment all computations have been done using an $18^4$ lattice and the standard Wilson plaquette action with coupling constant $\beta = 2.5$. This corresponds to lattice spacing $a \approx 0.073 \, \textrm{fm}$, when identifying $r_0$ with $0.46 \, \textrm{fm}$, which is close to the QCD value \cite{Philipsen:2013ysa}. For the computation of Wilson loops $W$ APE smeared spatial links have been used, while the links of the plaquettes $P_{0j}$ and $P_{kl}$ are unsmeared.


\section{\label{SEC004}Numerical results}
	
	\begin{figure}[b]
		\centerline{%
			\includegraphics[width=10cm]{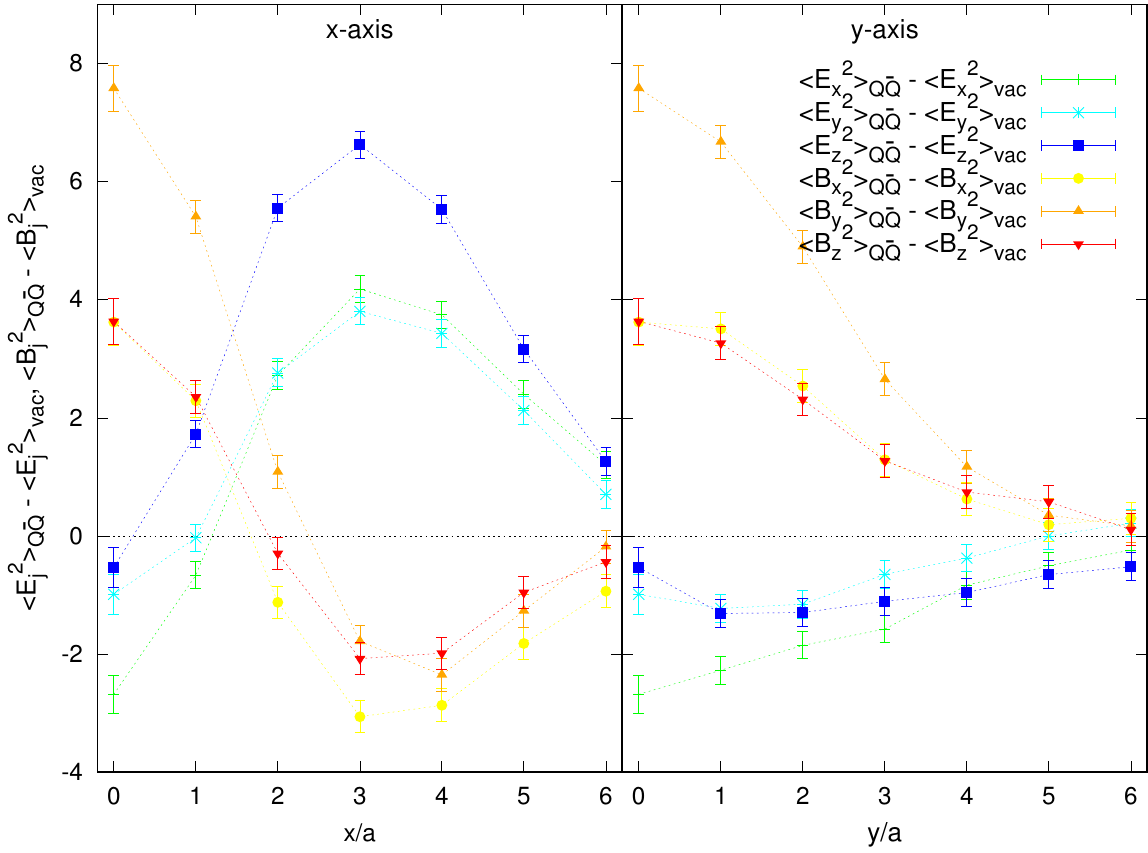}}
		\caption{$\braket{E_j(\textbf{x})^2}_{Q\bar{Q}} - \braket{E_j^2}_{\textrm{vac}}$ and $\braket{B_j(\textbf{x})^2}_{Q\bar{Q}} - \braket{B_j^2}_{\textrm{vac}}$ on the mediator axes for the hybrid static potential $\Pi_u$ for separation $r = 6a$.}
		\label{Fig:mediator_axes}
	\end{figure}

In Fig.\ \ref{Fig:separation_axis} we compare the field strengths of the ordinary static potential $\Sigma_g^+$ and the hybrid static potential $\Pi_u$ on the $Q \bar{Q}$ separation axis. In the vicinity of the color charges the field strengths are essentially the same for both potentials. However, a significant difference is observed in the region halfway between the quark and the antiquark. While the chromoelectric field strengths, in particular the $z$ component, i.e.\ the component parallel to the separation axis, dominate for $\Sigma_g^+$, they almost vanish in the hybrid case $\Pi_u$. At the same time, the chromomagnetic field strengths are large and positive for $\Pi_u$, while they even yield negative contributions to the energy density for the ordinary static potential $\Sigma_g^+$. 

In Fig.\ \ref{Fig:mediator_axes} we show again the field strengths of the hybrid static potential $\Pi_u$, this time along the two axes perpendicular to the separation axis originating in the center between the quark and the antiquark. These axes, which are parallel to the $x$ and the $y$ axes are also referred to as mediator axes. The field strengths along the two axes are not identical, i.e.\ the hybrid $\Pi_u$ state we excite with the spatial links shown in Fig.\ \ref{Fig:Wilson_loop_Pi_u} has a rotationally non-invariant gluon distribution. The general picture is the same as before: In the hybrid case $\Pi_u$ the chromomagnetic field strengths dominate, while the chromoelectric field strengths contribute in a negative way. Such a behavior is not surprising, since it is also possible and common to excite the $\Pi_u$ sector by locally inserting $\mathbf{r} \times \mathbf{B}$ halfway in both spatial parts of an ordinary Wilson loop (cf.\ e.g.\ the pNRQCD investigation \cite{Berwein:2015vca} or the recent lattice field theory study \cite{Wolf:2014tta}).

\begin{figure}[t]
		\centerline{%
			\includegraphics[width=8cm]{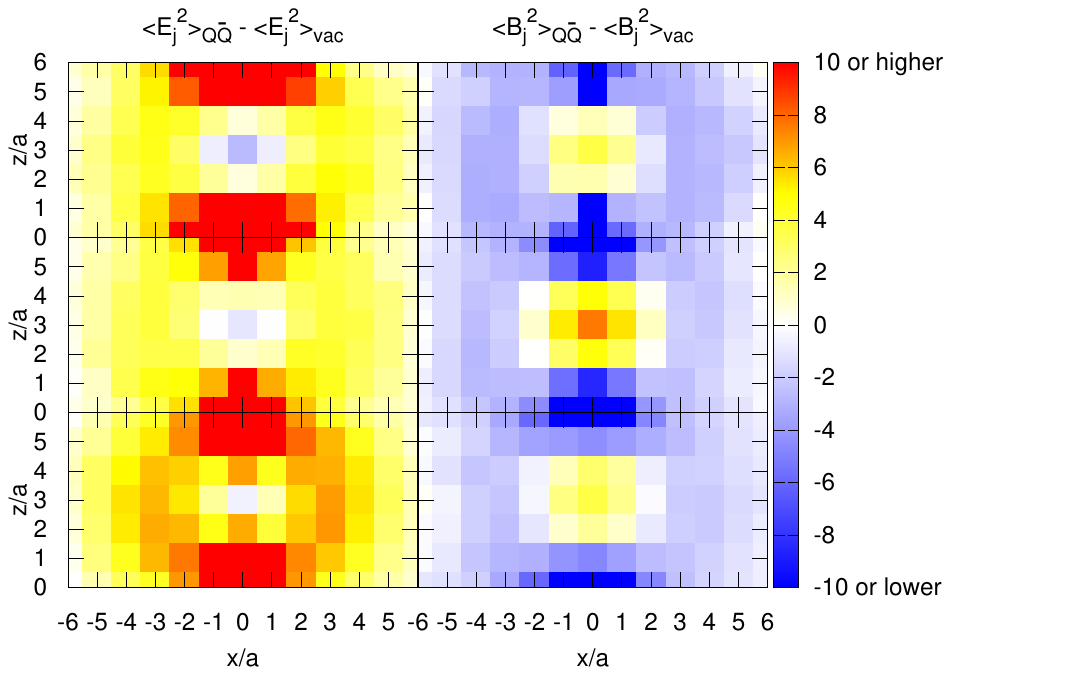}}
		\caption{$\braket{E_j(\textbf{x})^2}_{Q\bar{Q}} - \braket{E_j^2}_{\textrm{vac}}$ and $\braket{B_j(\textbf{x})^2}_{Q\bar{Q}} - \braket{B_j^2}_{\textrm{vac}}$ in the $x$-$z$ plane for the hybrid static potential $\Pi_u$ for separation $r = 6a$.}
		\label{Fig:heatmap}
	\end{figure}

Finally, in Fig.\ \ref{Fig:heatmap} we show the field strengths of the hybrid static potential $\Pi_u$ in the whole $x$-$z$ plane. Again, one can clearly see that chromomagnetic flux is localized halfway between the quark and the antiquark, while in the same region chromoelectric flux is strongly suppressed. We interpret this localized bump of chromomagnetic flux as gluons generating the hybrid quantum numbers of the meson.

It is also very interesting to compare our results to the results of the previously mentioned very recent and quite similar lattice field theory study \cite{Bicudo:2018yhk}, where the computation of flux tubes of hybrid static potentials has been carried out for the first time. The spatial parts of the Wilson loops used in \cite{Bicudo:2018yhk} to excite the $\Pi_u$ sector are somewhat different than in our work. While our two mass-degenerate $\Pi_u$ hybrid states have defined $\epsilon$, the two mass-degenerate $\Pi_u$ hybrid states in \cite{Bicudo:2018yhk} can be distinguished by the sign of the angular momentum, i.e.\ by $L_z = +1$ and $L_z = -1$. It is, however, easy to relate the two sets of states as well as their gluon distribution analytically: adding our field strengths in $x$, $y$ and $z$ direction corresponds to the rotationally invariant quantities computed in \cite{Bicudo:2018yhk}. Comparing e.g.\ our Fig.\ \ref{Fig:separation_axis} and \ref{Fig:mediator_axes} to Fig.\ 3, 4, 5 and 6 in \cite{Bicudo:2018yhk} shows, that there is a severe qualitative discrepancy between the results of \cite{Bicudo:2018yhk} and our results. While we observe a strong dominance of chromomagnetic flux near the center and of chromoelectric flux close to the color charges, no similar change from chromomagnetic to chromoelectric dominance has been found in \cite{Bicudo:2018yhk}. From a technical point of view the main difference between our work and \cite{Bicudo:2018yhk} is that we study gauge group SU(2), not SU(3), but it seems rather unlikely that the different gauge groups lead to flux tubes, which are drastically different. It will be very interesting to explore the reason for the observed discrepancy.


\section*{Acknowledgements}

L. M. thanks P. Bicudo and the other organizers of "Excited QCD 2018" for the invitation to give this talk. We thank C.\ Meyerdierks for important contributions at an early stage of this work \cite{Meyerdierks:2017}. We acknowledge helpful discussions with P.\ Bicudo, O.\ Philipsen and C.\ Reisinger. M.W.\ acknowledges support by the DFG (German Research Foundation), grant WA 3000/2-1. This work was supported in part by the Helmholtz International Center for FAIR within the framework of the LOEWE program launched by the State of Hessen.



\end{document}